\documentclass[%
 reprint,
 twocolumn,
 amsmath,amssymb,
 aps,
 prl,
 nofootinbib,
 tightenlines,
 nobibnotes
]{revtex4-2}

\usepackage{graphicx}% Include figure files
\usepackage{dcolumn}% Align table columns on decimal point
\usepackage{bm}% bold math
\usepackage{hyperref}% add hypertext capabilities
\usepackage{color}

\usepackage{gensymb}

\definecolor{Green}{RGB}{0,100,0}
\definecolor{Purple}{RGB}{102,0,255}
\definecolor{Blue}{RGB}{51,153,255}
\definecolor{Red}{RGB}{151,010,010}

\newcommand{\av}[1]{{\textcolor{Green}{#1}}}

\definecolor{Orange}{RGB}{255,69,0}

\begin{document}

\title{Tilted Poincar\'e Sphere Geodesics}

\author{Andrew A. Voitiv$^1$, Mark T. Lusk$^2$,}
\author{Mark E. Siemens$^1$}%
\email[]{msiemens@du.edu}%
\affiliation{$^1$Department of Physics and Astronomy, University of Denver, 2112 E. Wesley Avenue, Denver, CO 80208, USA
}%
\affiliation{
$^2$Department of Physics, Colorado School of Mines, 1500 Illinois Street, Golden, CO 80401, USA
}%

\date{\today}

\begin{abstract}
We provide the first experimental demonstration
of geometric phase generated in association with closed Poincar\'e Sphere trajectories comprised of geodesic arcs that do not start, end, or necessarily even include, the north and south poles that represent pure Laguerre-Gaussian modes. Arbitrarily tilted (elliptical) single vortex states are prepared with a spatial light modulator, and Poincar\'e Sphere circuits are driven by beam transit through a series of $\pi$-converters and Dove prisms.
\end{abstract}

\maketitle
When the Poincar\'e Sphere (PS) for Gaussian modes was introduced, the connection was immediately made between trajectories produced by linear optics and the accumulation of  geometric phase \cite{Padgett1999Poincare-sphereMomentum}. Within a closely related setting, the measurement of geometric phase had already been well-established by then for the polarization of light \cite{Simon1988EvolvingExperiment, Bhandari1988ObservationInterferometer, Kwiat1991ObservationPhoton}, as first predicted by Pancharatnam even before the context of Berry's seminal work \cite{Pancharatnam1956GeneralizedPencils, Berry1984QuantalChanges}. This parallel offered a convenient way of interpreting the underlying physics by comparing polarization states with their orbital angular momentum (OAM) counterparts: namely, circularly polarized states (Laguerre-Gaussian modes) on the poles, linearly polarized states (Hermite-Gaussian modes) on the equator, and mixtures of these states otherwise throughout the rest of the spherical surface \cite{Gutierrez-Cuevas2019GeneralizedVectors}. Likewise, the optics that generate trajectories on the PS are functionally analogous, such as quarter-waveplates ($\frac{\pi}{2}$-converters) and half-waveplates ($\pi$-converters). A series of $\frac{\pi}{2}$ mode converters and Dove prisms was used by Galvez et al. to transform beams from a fundamental Lageurre-Gaussian mode at a pole, to Hermite-Gaussian modes on the equator, and back to form closed circuits on the PS \cite{Galvez2003GeometricMomentum}. Using a collinear Gaussian beam for interferometric measurements, the authors extracted a geometric phase that scaled linearly with the rotation of the transforming optics---just as seen in the polarization counterpart many years before (e.g., \cite{Simon1988EvolvingExperiment}).

\begin{figure}[h!]
\centering
\includegraphics[width=.88\linewidth]{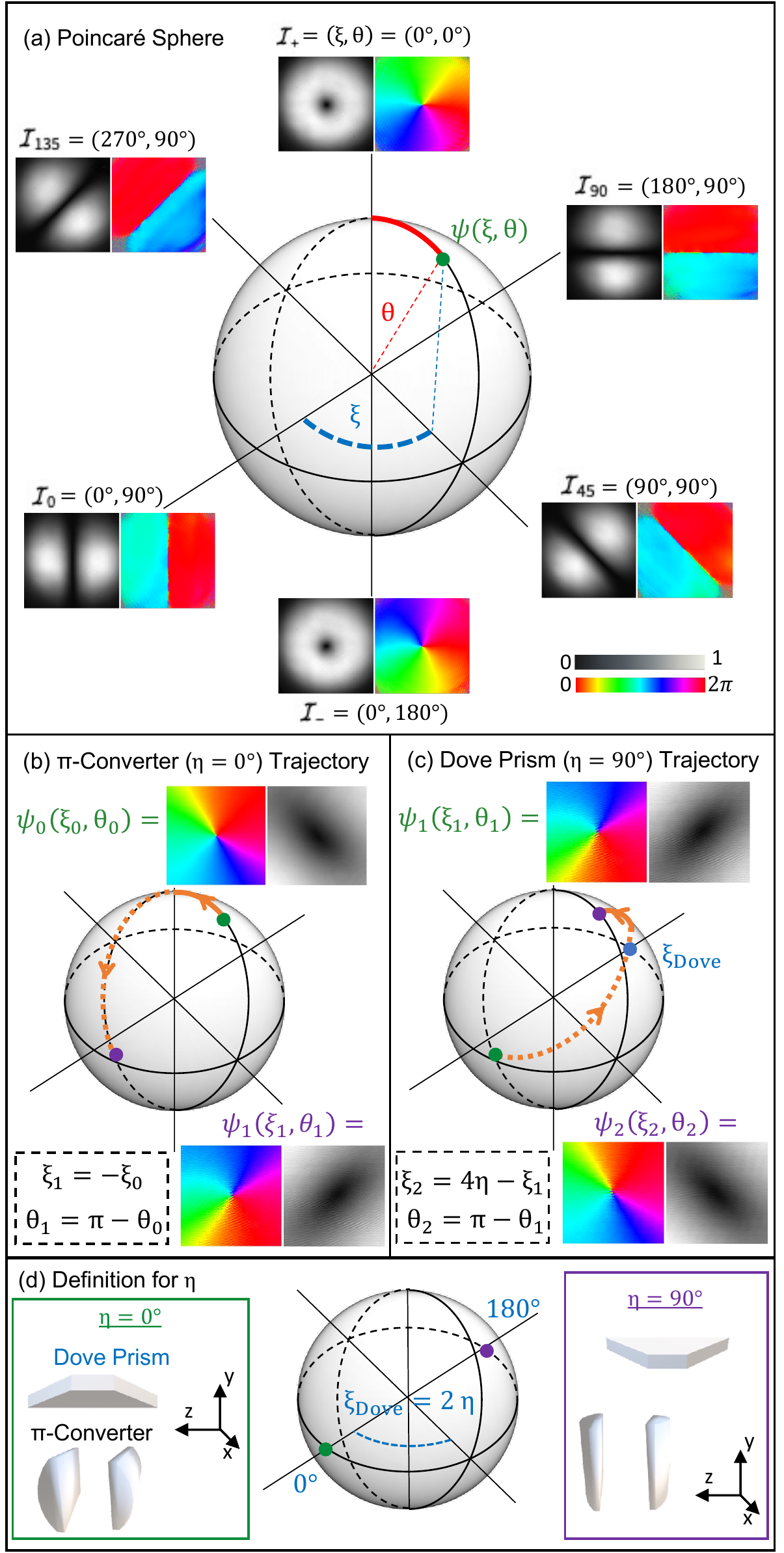}
\caption{(a) Vortex Poincar\'e Sphere (PS). Axes are labelled pictorially with experimental measurements of the mode amplitude (grayscale) and phase (hue), along with spherical coordinates $(\xi, \theta)$. North and south poles are also labelled with $\mathcal{I_{+}}$ and $\mathcal{I_{-}}$, respectively, for ``positively'' and ``negatively'' charged circular vortices. Similarly, the equatorial axes are labelled with $\mathcal{I}_{0}$, $\mathcal{I}_{45}$, $\mathcal{I}_{90}$, and $\mathcal{I}_{135}$, corresponding with the tilt-angle (in degrees) of the Hermite-Gaussian modes. An arbitrary field on the PS is represented by the green dot, $\psi(\xi,\theta)$. (b) Example geodesic trajectory of an arbitrary state, $\psi_{0}(90\degree, 37\degree)$, through a $\pi$-converter; the final state after the converter is $\psi_{1}(270\degree,143\degree)$, as described by Eqn. \ref{piconverterangles}. (c) Example geodesic trajectory of an initial state, $\psi_{1}(270\degree,143\degree)$, through a Dove prism. The Dove prism is rotated at $\eta = 90\degree$, ``sideways'' orientation, which picks a point on the equator $\xi_{\mathrm{Dove}} = 2 \eta$ to transform the $\xi$ angle about symmetrically. The $\xi_{1}=270\degree$ of $\psi_{1}$ is transformed to $\xi_{2}=90\degree$ passing through $\xi_{\mathrm{Dove}} = 180\degree$ on the PS. See Eqn. \ref{doveprismangles} for details. (d) Physical depictions of the orientation $\eta$ and its relation to $\xi_{\mathrm{Dove}}$ on the PS.}
\label{fig:ps}
\end{figure}

Consider the first-order Gaussian mode sphere of Fig. \ref{fig:ps}, which we will refer to as the ``vortex PS'' to emphasize that the beam waists and waist positions are not fixed on such spheres (particularly between start and end points through focusing optics). Previous experimentally considered trajectories on this PS have been confined to geodesic, or ``great circle,'' paths \cite{Courtial1998MeasurementMomentum, Padgett1999Poincare-sphereMomentum, Galvez2003GeometricMomentum}. Such trajectories are experimentally favorable because the total phase accumulation measured in the beam is strictly geometric---there is a lack of dynamic phase specifically associated with the mode transformations (distinct from propagation and oscillatory phase, also referred to as ``dynamic'') \cite{Bhandari1989SynthesisApproach, vanEnk1993GeometricTransfer}. While starting on the poles or equator is convenient, it is also possible to experimentally access arbitrary starting and ending points on the PS, something that has yet to be realized. This amounts to starting and ending with tilted vortices that have an elliptical core structure, which we anticipate will find applications in related developing fields of geometric phase accumulation, such as in nonlinear optics, which generally have used starting points only on the poles \cite{Karnieli2022TheConversion}.

%%% Figure One %%%
%%% (a) Data from: "PS images for paper.nb" from Small Circles project %%%
%%% (b) & (c) Data from the Great Circles "Images of Data"

In this Letter, we demonstrate arbitrary geodesic circuits on a vortex PS utilizing common optical components: cylindrical lenses, Dove prisms, and a spatial light modulator (SLM). The SLM is used to construct vortex beams that are ``virtually tilted,'' allowing for controlled initial positions on the PS off of the poles and equator. We demonstrate how geodesic trajectories, and their accumulated geometric phases, can be obtained for arbitrary initial states of vortices. Our measurements are consistent with the fact that the geometric phase is invariant for arbitrary initial locations on the PS for a given set of transforming optics yielding geodesic arcs.

\begin{figure}[h]
\centering
\includegraphics[width=.88\linewidth]{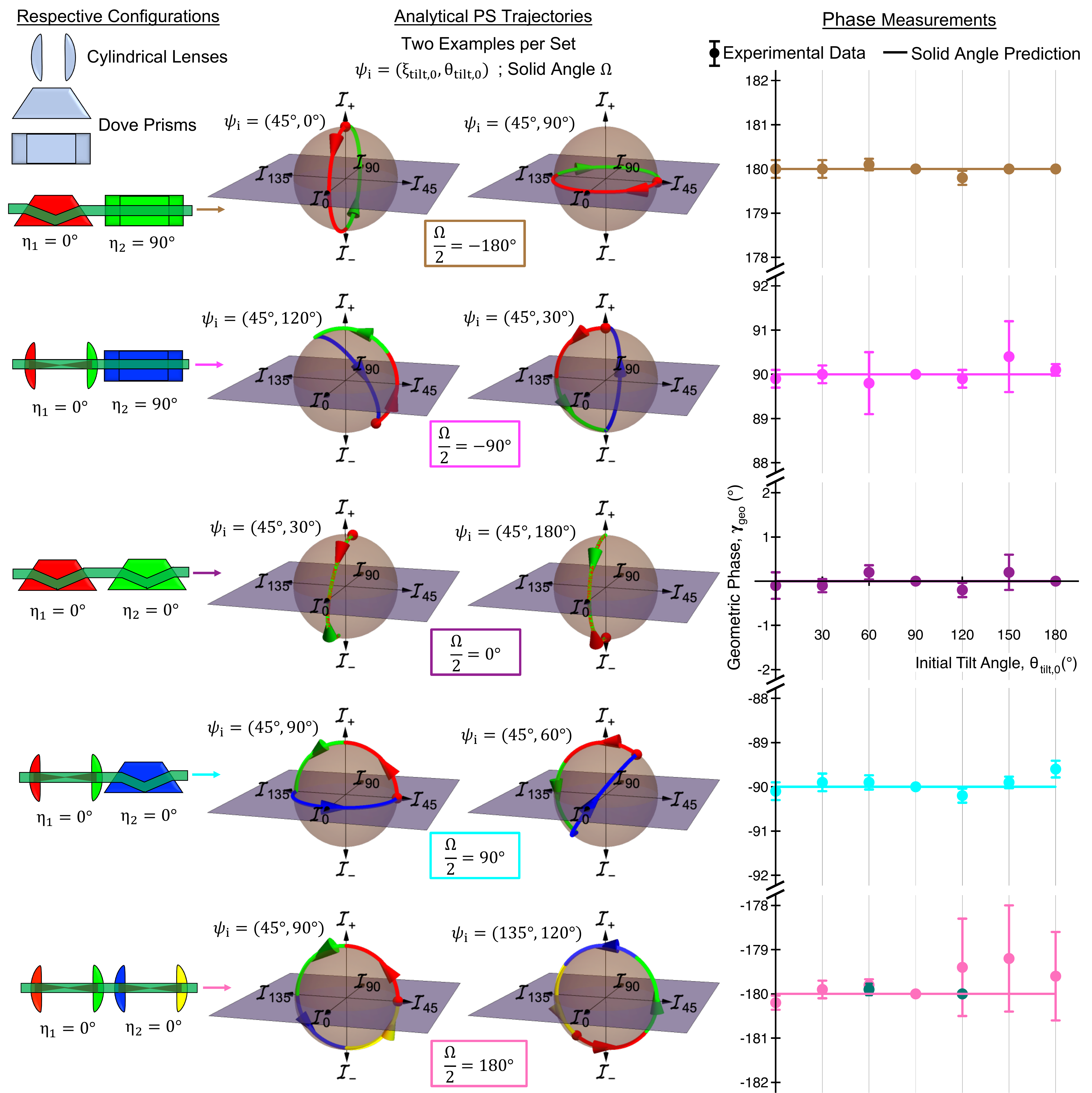}
\caption{(Left Column) Configurations of $\pi$-converters and Dove prisms used to measure different settings of geodesic geometric phase. Each optic is colored to match its associated transformation trajectory on the analytical plots at right. Relative rotations of the two pairs of optics are denoted by $\eta_1$ and $\eta_2$, respectively, as defined in Fig. \ref{fig:ps} (d). (Middle Column) Analytical circuits of vortex transformations on a vortex PS \cite{Lusk2021TheLens}. There are two examples (two different initial conditions, $\psi_{\mathrm{i}}$) for each set of optics used. The half solid angle enclosed by each loop (equivalent for a given set of optics) is labeled between the two examples. Each optic (each cylindrical lens and Dove prism) transforms the vortex independently, and therefore each optic's corresponding trajectory on the PS is denoted by a different color. (Right Column) Experimental measurements of the geometric phase (vertical axis) as a function of initial starting locations on the sphere in terms of initial tilt-angles, $\theta_{0,\mathrm{tilt}}$ (horizontal axis). Data-point colors correspond to different arrangements of $\pi$-converters and Dove prisms. There are two \av{teal} data points on the row for $\gamma_{\mathrm{geo}}=-180\degree$, which have $\xi_{\mathrm{tilt},0} = 135\degree$ (rather than $45\degree$ like all other points), which follow the same geodesic circuit as the other points in this row (as depicted on the spheres in the middle column).Conversions from the ``virtual tilt'' angles \cite{Andersen2021HydrodynamicsFluids} to PS coordinates of the paper are: $\theta = \arccos{\biggl[ \frac{4\cos{(\theta_{\mathrm{tilt}})} }{3 + \cos{(2\theta_{\mathrm{tilt}})}} \biggl]}$ and $\xi = 2 \, \xi_{\mathrm{tilt}}$.}
\label{fig:behemoth}
\end{figure}

To construct arbitrary vortex states on the PS, use the coordinates of Fig. \ref{fig:ps} and a basis in terms of first-order ($\pm 1$) Laguerre-Gaussian (LG) modes. Then, a natural way to write a given field is
\begin{equation} \label{natural}
    \psi(\xi,\theta) = \frac{1}{\sqrt{2}} \biggl[ \cos{\left( \frac{\theta}{2} \right)} e^{+i \xi} \mathrm{LG}_{\mathcal{I}_{+}} + \sin{\left( \frac{\theta}{2} \right)} e^{-i \xi} \mathrm{LG}_{\mathcal{I}_{-}} \biggl],
\end{equation}
where the polar angle $\theta$ tells us about the ratio of the two fundamental modes (on the poles) and the azimuthal angle $\xi$ shows the relative phase between them, for making any mode on the PS. In this experiment, however, we chose to employ the perspective of ``virtual tilt,''\cite{Andersen2021HydrodynamicsFluids} which works by taking a vortex on the north pole ($\mathcal{I}_{+}$) and stretching it by angle $\theta_{\mathrm{tilt}}$ and rotating it about its centroid by $\xi_{\mathrm{tilt}}$. For clarity, we focus the discussion on only the PS coordinates of Eqn. \ref{natural}. See the caption of Fig. \ref{fig:behemoth} for relations between the two sets of coordinates.

%%% Figure Two %%%
\begin{figure}[h!]
\centering
\includegraphics[width=\linewidth]{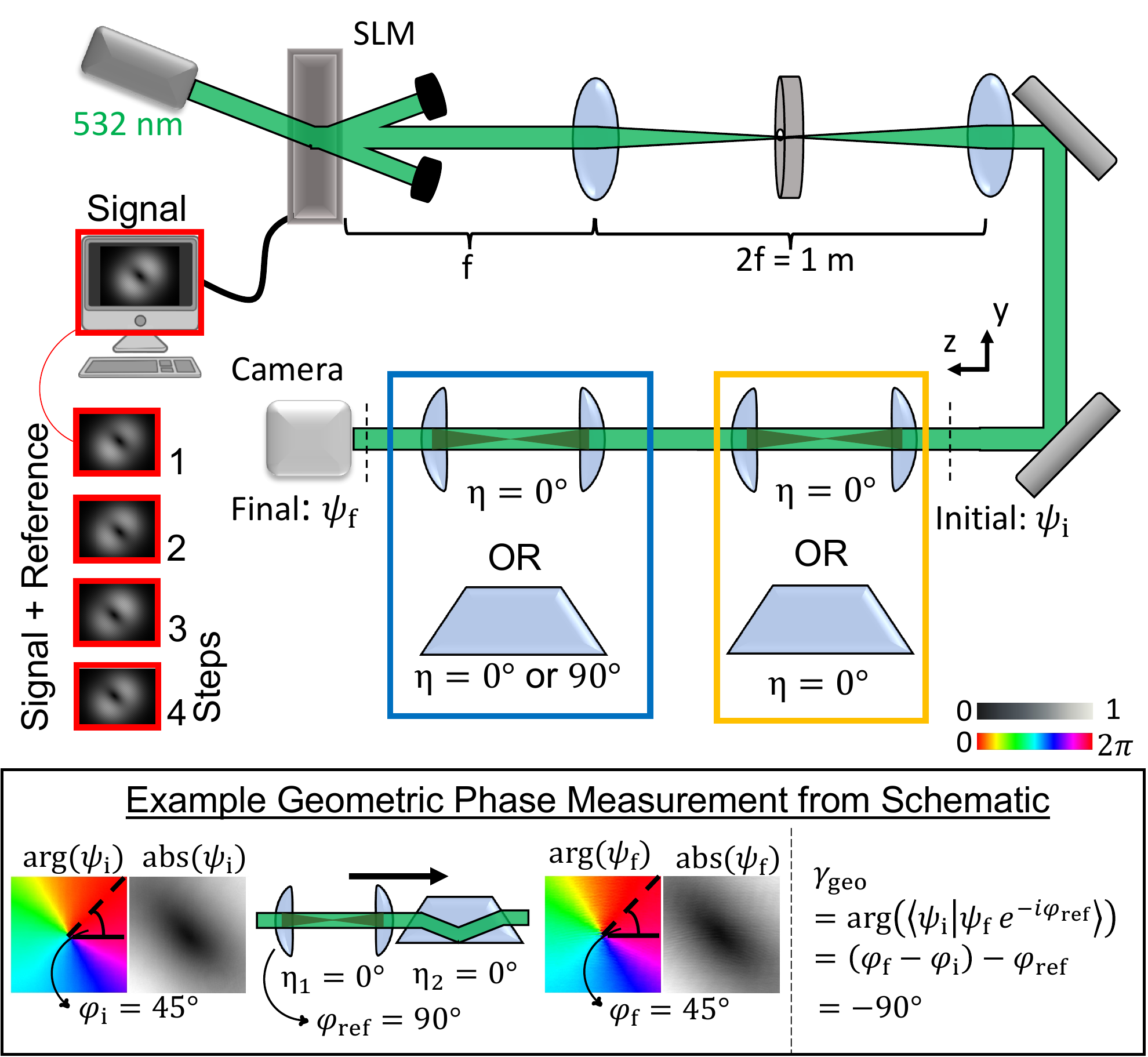}
\caption{ \textit{Schematic.} A $\lambda = 532$ nm, collimated, single-mode Gaussian passes through a hologram on a transmission-SLM, from which the first diffracted order is imaged onto the different possible pairs of $\pi$-converters and Dove prisms. The SLM is controlled by an Epson 83H projector \cite{Huang2012ALabs}, and an Andor Zyla sCMOS detector captures transverse images of the vortex beams. Boxed in red are the gratings used on the SLM, including four with a superposed collinear Gaussian reference for phase measurements under different reference phase steps \cite{Andersen2019CharacterizingHolography}. Boxed in yellow are the first series of optics encountered by the beam, to bring the vortex from its initial state on the PS to some intermediate state. Boxed in blue are the second series of optics used to bring the intermediate state back to the starting location on the PS, to form closed geodesic circuits. Cylindrical lenses are separated by two times their focal length and are both locked to focus in the $y$-axis ($\eta = 0\degree$). Dove prisms are either oriented along the same $y$-axis, or are rotated relatively by $\eta=90\degree$. \textit{Inset.} One example of experimental measurements of the complex field of initial and final vortex states, $\psi_{\textrm{i}}$ and $\psi_{\textrm{f}}$. The rotation of the ``phase $=0 \, (2\pi)$" line (red color, $\varphi$ labels) is ultimately how the geometric phase is measured, similarly to tracking the rotation of interferograms \cite{Galvez2003GeometricMomentum}.}
\label{fig:schematic}
\end{figure}
%%% Grating image is: "grating_45_60_amp" %%%

A vortex of the form of Eqn. \ref{natural} can be constructed experimentally by programming a hologram to impart the proper phase and amplitude structure onto an incoming Gaussian laser beam, as shown with a transmission-SLM in Fig. \ref{fig:schematic}. The hologram of the desired mode is a plane wave grating that has the structural form
\begin{equation}
    \mathrm{Hologram}(x,y) = \mathrm{Abs}[e^{i \mathrm{Arg}[\psi]} + e^{i k_{\mathrm{g}} x}] \times \frac{\mathrm{Abs}[\psi]}{\mathrm{Max}[\mathrm{Abs}[\psi]]},
\end{equation}
for $\psi$ of Eqn. \ref{natural}, and wave number $k_{\mathrm{g}}$ to set the grating spacing. From generation at the SLM, the vortex mode is then imaged onto a series of $\pi$-converters and Dove prisms.

%%% Figure Three %%%
\onecolumngrid

\twocolumngrid
%%% Source of Analytical Trajectories: PS_Trajectories_Great_Circles_Paper_120221_Figures_for_Drew-4.nb
%%% Tabled Data: Geometric Phase > Great Circles > Paper
%%% Raw Data (lab): Geometric Phase > Great Circles    

The theory we employ to predict the trajectories made on the PS in our experiments is based on tracking the vortex transformation as a function of propagation \textit{through} the optical lenses \cite{Lusk2021TheLens} (or Dove prisms). This theory is used to plot the PS trajectories seen in the middle column of Fig. \ref{fig:behemoth}, which can be used to determine a solid angle enclosed by the transformations.

In order to enforce geodesic trajectories for an arbitrary initial vortex state, one must choose a configuration of $\pi$-converters/Dove prisms that will ``maximally change'' the vortex state on the PS, following the analogy with half-waveplates and polarization states \cite{Padgett1999Poincare-sphereMomentum}. In Fig. \ref{fig:ps} (b), one notices that keeping the $\pi$-converter locked to $\eta = 0\degree$ will transform a vortex state from $\xi = 90\degree$ (``diagonally-oriented'') to $\xi = 270\degree$ (oppositely ``diagonally-oriented''). To move on a geodesic, the relative angle between optics and vortex orientations must be: $\eta_{\mathrm{relative}} = |\eta - \frac{1}{2} \, \xi| = 45\degree$, for $\eta$ and $\xi$ defined in Fig. \ref{fig:ps}. A $\pi$-converter transforms a vortex mode from an initial position $(\xi_{\mathrm{i}},\theta_{\mathrm{i}})$ to a final position according to
\begin{equation} \label{piconverterangles}
    \pi \textrm{-converter: } \xi_{\mathrm{f}} = -\xi_{\mathrm{i}}, \hspace{.1in} \theta_{\mathrm{f}} = 180\degree - \theta_{\mathrm{i}},
\end{equation}
which is constrained to a $180\degree$ arc that goes through at least one pole of the PS, as depicted in Fig. \ref{fig:ps} (b), because of the geodesic-\textit{only} restriction.

A Dove prism transforms a vortex by reflecting it about a specific point on the equator of the PS. This produces a geodesic trace on the PS that does not need to encompass one of the poles. If $\eta_{\mathrm{relative}} = 45\degree$, then an intermediate state will transform to a final state that coincides with the initial state on the PS. In particular, the orientation of the Dove prism, $0\degree \le \eta \le 180\degree$, is identified with a point on the PS equator ($\xi_{\mathrm{Dove}} = 2 \eta$ as shown in Fig. \ref{fig:ps} (d)) and the mode is reflected about this point. Polar angle, $\theta_{\mathrm{i}}$, is thus transformed to $\theta_{\mathrm{f}}=180\degree-\theta_{\mathrm{i}}$ (flipping the topological charge of the vortex). The associated PS trajectory always crosses the equator at $\xi_{\mathrm{Dove}}$. This Dove reflection results in total angle of $2 \xi_{\mathrm{Dove}}$ between initial and final azimuthal orientations. A Dove prism trajectory is depicted in Fig. \ref{fig:ps} (c), in which to return to the same initial state shown in (b) one must choose either $\eta=0\degree$ or $90\degree$. Considering both PS angles, the Dove prism transforms a vortex mode according to:
\begin{equation} \label{doveprismangles}
     \textrm{Dove prism: } \xi_{\mathrm{f}} = 4\eta - \xi_{\mathrm{i}}, \hspace{.1in} \theta_{\mathrm{f}} = 180\degree - \theta_{\mathrm{i}}.
\end{equation}
Example PS trajectories described by these equations are shown in the middle column of Fig. \ref{fig:behemoth}. The geometric phase is equal to (minus) half the solid angle made by a complete circuit.

Once a closed circuit on the PS has been completed by transforming a vortex mode through any of the series of optics used, one can measure the geometric phase accumulated during the process. This is done fundamentally by comparing the final state to the initial state, as depicted in the inset of Fig. \ref{fig:schematic}. One can take the argument of the ``overlap'' between two complex fields if circuits are confined to geodesic arcs: $\gamma_{\mathrm{geo}} = \mathrm{Arg}[ \langle \psi_{\mathrm{i}} | \psi_{\mathrm{f}} \rangle ]$ \cite{Pancharatnam1956GeneralizedPencils, vanEnk1993GeometricTransfer}. It is also possible to track the rotation of interferograms, after different PS trajectories, to compare with the initial state interferogram \cite{Galvez2003GeometricMomentum}. Here, we employ a combination of the two approaches. We measure phase maps of a vortex beam using a collinear Gaussian reference beam with the technique of phase-shifting digital holography \cite{Andersen2019CharacterizingHolography}; these measurements are the typical hue maps of $\mathrm{arg}[\psi]$ that depict the swirling phase gradients of the vortex. These maps contain information of where the phase gradients start from zero, $\mathrm{arg}[\psi] = 0$. We call this angle $\varphi$ and it is measured with respect to the positive x-axis of the transverse plane of the mode.

Lastly, we measure the geometric phase between initial and final states by taking a pseudo-overlap between the states with these measurements: $\gamma_{\mathrm{geo}} = \mathrm{arg}[\psi_{\mathrm{f}}] - \mathrm{arg}[\psi_{\mathrm{i}}] = \varphi_{\mathrm{f}} - \varphi_{\mathrm{i}}$. To report the data in the right column of Fig. \ref{fig:behemoth} quantitatively, we fit the measured phase maps to a model of the virtually tilted vortex \cite{Andersen2021HydrodynamicsFluids}, which returns the angle $\xi_{\mathrm{tilt}}$ as a fitted parameter, which is identical to $\varphi$. Each data point of Fig. \ref{fig:behemoth} is the average of five measurements for the same set-up and initial tilt angles. The error bars associated with each point were calculated by taking one standard deviation of the measurements of each $\varphi$ and then propagating the deviations together to be the error of the average \cite{taylor1997introduction}. The errors arise from typical experimental beam drift between camera acquisitions and are insignificant on the scale of matching the entire set of theoretical predictions. 

The Gaussian reference beam used in the construction of the phase maps is collinear with the vortex mode throughout the entire circuit. Because the Gaussian itself picks up a phase factor of $\frac{\pi}{2}$ through a $\pi$-converter \cite{Beijersbergen1993AstigmaticMomentum}, any phase accumulation of $\frac{\pi}{2}$ picked up by the vortex mode through a $\pi$-converter will not be measured because the phase measurements are all \textit{in reference} to that Gaussian beam. For this reason, we subtract $90\degree$ from the measurements of $\varphi_{\mathrm{f}} - \varphi_{\mathrm{f}}$ for each $\pi$-converter present in a circuit, to calculate the correct geometric phase accumulated by the vortex. (The Gouy phase of the vortex mode through the $\pi$-converters \cite{Beijersbergen1993AstigmaticMomentum} is purely geometric for geodesic circuits.) For Dove prisms, there is no phase accumulation for a zero-order Gaussian beam. With these corrections, the measured change in phase, $\varphi_{\mathrm{f}} - \varphi_{\mathrm{i}}$, is consistent with the geometric phase predicted by the enclosed solid angle from a closed loop on the PS. The results show that for a given geodesic circuit, the geometric phase is invariant towards changing the initial location of the vortex on the PS or even from reorienting the circuit around different areas of the sphere.

In conclusion, we have demonstrated the measurement of vortex transformations along a PS with arbitrary vortex initial states, and we have shown how to experimentally measure the geometric phase associated with a closed circuit comprised of geodesic arcs taken. This work is the first experimental realization, to the best of our knowledge, of geodesic geometric phase measured from arbitrary initial and final states on the PS.

\noindent \textbf{Funding.} W. M. Keck Foundation; NSF (1553905)

\noindent \textbf{Disclosures.} The authors declare no conflicts of interest.

\noindent \textbf{Data availability.} Data underlying all results presented are available from the authors upon reasonable request.

% Bibliography
\bibliography{Refs}

\end{document}